\documentclass{article}
\usepackage{amsmath}
\usepackage{amsfonts}
\usepackage{amssymb}
\usepackage{cmmib57}
\begin{document}
\begin{quotation}

\textsc{A Finslerian Cosmological Metric And Its Riemannian Kaluza-Klein Extension}
\vspace{0.11in}

\bigskip

{\huge \vspace{0.05in}\allowbreak }
\textsc{M. Ar\i k and D. \c{C}ift\c{c}i}

\allowbreak \medskip {\small Bo\u{g}azi\c{c}i University,
Department of Physics, Istanbul, Turkey}

\ \hspace{0.3in}

{\small \textbf{Abstract}. In this study a rotationally and
translationally invariant metric in Finsler space is investigated.
We choose to rewrite the metric in Riemanian space by increasing
the dimension of space-time and introducing additional coordinates
such that for specific values of these coordinates, the geodesics
of the four dimensional Finslerian space-time and six dimensional
Riemanian space-time are identical. Cosmological solutions
described by this metric give rise to an equation of state
corresponding to a space dominated by domain walls and an internal
space dominated by strings. \vspace{0.2in}}
\end{quotation}

The cosmological principle states that at each epoch the universe
has the same aspect from every point except for local
irregularities, i.e. the large-scale universe is homogeneous and
isotropic. The Robertson-Walker-Friedman (RWF) metric \cite[2,
3]{linde}, is thus accepted as the standard metric for such a
cosmology. A natural question is whether the RWF metric is unique
for a homogenous and isotropic cosmology. For a Riemannian metric
in $3+1$ dimensions the answer to this question is in the
affirmative. In the context of Finsler \cite[5]{finsler} geometry,
however, there may exist various homogenous and isotropic
metrics.\ Although Finsler geometry is mathematically well
investigated \cite{rund} an acceptable theory of Finslerian
general relativity does not exist. In this paper we investigate a
Finslerian metric which is both rotationally and
translationally symmetric. The symmetry is given by the six parameter group $%
ISO(3)$ acting on the $3$ dimensional Euclidean space $R^{3}$.
Choosing a point as origin in $R^{3}$, three of these parameters
correspond to rotations around this origin whereas the other three
parameters are translations. Such a metric is given by
\begin{equation}
ds^{2}=-\left( dt+b\left( t\right)
\sqrt{dx^{2}+dy^{2}+dz^{2}}\right) ^{2}+a^{2}\left( t\right)
\left( dx^{2}+dy^{2}+dz^{2}\right) ,  \label{1}
\end{equation}
where $a\left( t\right) $ and $b\left( t\right) $ are cosmic scale
factors. This metric has some similarities to the G\"{o}del metric
\cite[8]{godel}, which describes a rotating universe, to the four
dimensional Taub-Nut metric for $m>0,$ which is interpreted as a
gravitational monopole \cite{taub1}, and also to the
Gibbons-Hawking metric \cite[11]{gibbons}, which describes
gravitational instantons.

The metric given by (\ref{1}) is difficult to handle. Moreover a
Finslerian cosmological model is yet nonexistent. By introducing
two additional dimensions and a constant $\varepsilon $, the
geodesics of the four dimensional Finslerian space-time and six
dimensional Riemannian space-time with the following metric can be
made identical in the limit $\varepsilon \rightarrow 0.$ This
Riemannian metric is written in the form
\begin{equation}
ds^{2}=-\left( dt+b\left( t\right) d\sigma \right) ^{2}+\left(
a^{2}\left( t\right) -\lambda ^{2}\right) \left(
dx^{2}+dy^{2}+dz^{2}\right) +\lambda ^{2}d\sigma ^{2}+\varepsilon
^{2}d\lambda ^{2},  \label{2}
\end{equation}%
where $\lambda $ and $\sigma $ are new coordinates. This space is
six dimensional and it is invariant under the $ISO\left( 3\right)
$ group. The geodesics of the metric (\ref{1}) and (\ref{2}) can
be readily found by the variational principle
\begin{equation}
\delta \int ds=0.
\end{equation}%
which gives the geodesic equations as the Euler-Lagrange equations
of a Lagrangian. This Lagrangian for the Finslerian metric
(\ref{1}) is
\begin{equation}
L=\left[ -\left( t^{\prime }+b\left( t\right) \sqrt{x^{\prime
2}+y^{\prime 2}+z^{\prime 2}}\right) ^{2}+a^{2}\left( t\right)
\left( x^{\prime 2}+y^{\prime 2}+z^{\prime 2}\right) ^{2}\right]
^{1/2}.  \label{eqs3}
\end{equation}%
By the same consideration the Lagrangian for the Riemaniann metric is%
\begin{equation}
L=\left[ -\left( t^{\prime }+b\left( t\right) \sigma ^{\prime
}\right) ^{2}+\left( a^{2}\left( t\right) -\lambda ^{2}\right)
\left( x^{\prime 2}+y^{\prime 2}+z^{\prime 2}\right) +\lambda
^{2}\sigma ^{\prime 2}+\varepsilon ^{2}\lambda ^{\prime 2}\right]
^{1/2}  \label{eqs2}
\end{equation}%
where $q_{i}^{\prime }=dq_{i}/ds.$ To obtain the geodesics of
these actions
we use the Euler-Lagrange equations%
\begin{equation}
\frac{\partial L}{\partial q_{i}}=\frac{d}{ds}\left( \frac{\partial L}{%
\partial q_{i}^{\prime }}\right) .
\end{equation}%
First we calculate this for $q_{i}=\lambda $ from (\ref{eqs2}):%
\begin{equation}
-\lambda \left( x^{\prime 2}+y^{\prime 2}+z^{\prime 2}\right)
+\lambda \sigma ^{\prime 2}=\varepsilon \lambda ^{\prime }
\end{equation}%
the right hand side of which goes to zero since $\varepsilon
\rightarrow 0.$ For the coordinate $t$ in (\ref{eqs3}) we obtain
the Euler-Lagrange equation%
\begin{eqnarray}
&-&\!\!\!\!\dot{b}\sqrt{x^{\prime 2}+y^{\prime 2}+z^{\prime 2}}%
\left( t^{\prime }+b \sqrt{x^{\prime 2}+y^{\prime 2}+z^{\prime
2}}\right)+a\dot{a}\left( x^{\prime 2}+y^{\prime
2}+z^{\prime 2}\right) ^{2}=  \notag \\
&-&\!\!\!\frac{d}{ds}\left( t^{\prime }+b \sqrt{x^{\prime
2}+y^{\prime 2}+z^{\prime 2}},\right)
\end{eqnarray}%
whereas for the $t$ coordinate in (\ref{eqs2}):%
\begin{equation}
-\dot{b}\, \sigma ^{\prime }\left( t^{\prime
}+b\left( t\right) \sigma ^{\prime }\right) +a\dot{a}\left( x^{\prime 2}+y^{\prime 2}+z^{\prime
2}\right) =\frac{d}{ds}\left[ -\left( t^{\prime }+b\sigma ^{\prime }\right) \right],
\end{equation}%
where $a\left( t\right)=a$ and $b\left( t\right)=b$. Last two equations indicate that
\begin{equation}
\sigma ^{\prime 2}=x^{\prime 2}+y^{\prime 2}+z^{\prime 2}.
\label{eqs4}
\end{equation}%
hence we can substitute $d\sigma =\sqrt{dx^{2}+dy^{2}+dz^{2}}$ in
(\ref{2}) upon which (\ref{1}) and (\ref{2}) become partly
identical. Now we integrate
the $x,y,z$ components of the Euler-Lagrange equations obtained from (\ref%
{eqs3}) and using (\ref{eqs4}) we find
\begin{equation}
\left(a^{2}-b^{2}-\frac{b\,
t^{\prime }}{\sigma ^{\prime }}\right) \mathbf{X}^{\prime
}=\mathbf{C} \label{eq5}
\end{equation}%
The corresponding equations for the geodesics obtained from
(\ref{eqs2}) can be integrated to yield
\begin{equation}
\left( a^{2}-\lambda ^{2}\right) \mathbf{X}^{\prime }=%
\mathbf{D}  \label{eq6}
\end{equation}%
If we perform the same calculation for the other $\sigma $ component in (\ref%
{eqs2}) we obtain%
\begin{equation}
-b\left( t^{\prime }+b\, \sigma
^{\prime }\right) +\lambda ^{2}\sigma ^{\prime }=A
\end{equation}%
where $\mathbf{C},\mathbf{D},A$ are constants of integrations. If
we solve
the last equation for $\lambda $ we find%
\begin{equation}
\lambda ^{2}=\frac{1}{\sigma ^{\prime }}\left[b\, t^{\prime }+A%
\right]+b^{2} .
\end{equation}%
Substituting this in (\ref{eq6}) gives
\begin{equation}
\left[ a^{2}-b^{2}-\frac{1}{\sigma ^{\prime }}\left(b\,
t^{\prime }+A\right)\right]
\mathbf{X}^{\prime }=\mathbf{D.} \label{eq7}
\end{equation}%
From (\ref{eq5}) and (\ref{eq7}) we obtain that%
\begin{equation}
\frac{A}{\sigma ^{\prime }}=C_{x}-D_{x}=C_{y}-D_{y}=C_{z}-D_{z}.
\end{equation}%
Since $\sigma ^{\prime }$ is variable the only acceptable solution
of this equation is $A=0,$ $\mathbf{C=D.}$ We have shown that for
$A=0$ geodesics of (\ref{1}) and geodesics of (\ref{2}) are
identical.

We now calculate the Riemannian curvature of the metric (\ref{2})
using the Cartan equations of structure in an orthonormal basis.
We choose the orthonormal basis one forms
\begin{eqnarray}
e^{4}&=&i\left( dt+b\left( t\right) d\sigma \right)\\
e^{1}&=&\left(a^{2}-\lambda ^{2}\right) ^{1/2}dx \\
e^{2}&=&\left( a^{2}-\lambda^{2}\right) ^{1/2}dy \\
e^{3}&=&\left( a^{2}-\lambda ^{2}\right)^{1/2}dz\\
e^{5}&=&\lambda d\sigma\\
e^{6}&=&\varepsilon d\lambda .\\
\end{eqnarray}%
The non-zero components of the Einstein tensor are found as
\begin{eqnarray}
G_{11}=G_{22}=G_{33}&=&\frac{a^{2}\dot{a}^{2}}{\left(
a^{2}-\lambda ^{2}\right)
^{2}}-\frac{a^{2}\dot{a}^{2}b^{2}}{\lambda ^{2}\left(
a^{2}-\lambda ^{2}\right) ^{2}}-\frac{\lambda ^{2}}{\varepsilon
^{2}\left(
a^{2}-\lambda ^{2}\right) ^{2}}  \notag \\
&-&\!\!\!\frac{2\dot{a}^{2}}{\left( a^{2}-\lambda ^{2}\right) }-\frac{2a\ddot{a}}{%
\left( a^{2}-\lambda ^{2}\right)
}+\frac{2b^{2}\dot{a}^{2}}{\lambda ^{2}\left( a^{2}-\lambda
^{2}\right) }+\frac{2b^{2}a\ddot{a}}{\lambda
^{2}\left( a^{2}-\lambda ^{2}\right) }\notag \\
&+&\!\!\!\frac{4a\dot{a}b\dot{b}}{\lambda ^{2}\left( a^{2}-\lambda ^{2}\right) }-%
\frac{4}{\varepsilon ^{2}\left( a^{2}-\lambda ^{2}\right) }+\frac{b\ddot{b}}{%
\lambda ^{2}}+\frac{\dot{b}^{2}}{\lambda ^{2}}
\end{eqnarray}

\begin{eqnarray}
G_{44} &=&-\frac{3a^{2}\dot{a}^{2}}{\left( a^{2}-\lambda ^{2}\right) ^{2}}+%
\frac{3a\dot{a}b\dot{b}}{\lambda ^{2}\left( a^{2}-\lambda ^{2}\right) }+%
\frac{3\dot{a}^{2}b^{2}}{\lambda ^{2}\left( a^{2}-\lambda ^{2}\right) } \notag\\
&+&\frac{3a\ddot{a}b^{2}}{\lambda ^{2}\left( a^{2}-\lambda ^{2}\right) }-%
\frac{6}{\varepsilon ^{2}\left( a^{2}-\lambda ^{2}\right) }\\
G_{55} &=&-\frac{3\dot{a}^{2}}{\left( a^{2}-\lambda ^{2}\right) }+\frac{3a%
\ddot{a}}{\left( a^{2}-\lambda ^{2}\right) }+\frac{3a\dot{a}b\dot{b}}{%
\lambda ^{2}\left( a^{2}-\lambda ^{2}\right) }\notag \\
&+&\frac{3a^{2}\dot{a}^{2}b^{2}}{\lambda ^{2}\left( a^{2}-\lambda
^{2}\right) }-\frac{3}{\varepsilon ^{2}\left( a^{2}-\lambda
^{2}\right) }\\
G_{66} &=&-\frac{3\dot{a}^{2}}{\left( a^{2}-\lambda ^{2}\right) }+\frac{3a%
\ddot{a}}{\left( a^{2}-\lambda ^{2}\right) }+\frac{6a\dot{a}b\dot{b}}{%
\lambda ^{2}\left( a^{2}-\lambda ^{2}\right) }  \notag \\
&+&\!\!\frac{3\dot{a}^{2}b^{2}}{\lambda ^{2}\left( a^{2}-\lambda ^{2}\right) }+%
\frac{3a\ddot{a}b^{2}}{\lambda ^{2}\left( a^{2}-\lambda ^{2}\right) }+\frac{%
\lambda ^{2}}{\varepsilon ^{2}\left( a^{2}-\lambda ^{2}\right) }\notag \\
&-&\!\!\frac{3}{\varepsilon ^{2}\left( a^{2}-\lambda ^{2}\right) }+\frac{b\ddot{b%
}}{\lambda ^{2}}+\frac{\dot{b}^{2}}{\lambda ^{2}}\\
G_{45}&=&-\frac{3i\dot{a}^{2}b}{\lambda \left( a^{2}-\lambda ^{2}\right) }-%
\frac{3ia\ddot{a}b}{\lambda \left( a^{2}-\lambda ^{2}\right) }+\frac{3ia^{2}%
\dot{a}^{2}b}{\lambda \left( a^{2}-\lambda ^{2}\right) ^{2}}\\
G_{46}&=&\frac{3ia\dot{a}\lambda }{\varepsilon \left( a^{2}-\lambda
^{2}\right) ^{2}}\\
G_{56}&=&\frac{6a\dot{a}b}{\varepsilon \left( a^{2}-\lambda ^{2}\right) ^{2}}-%
\frac{3a^{3}\dot{a}b}{\lambda ^{2}\varepsilon \left( a^{2}-\lambda
^{2}\right) ^{2}}-\frac{\dot{b}}{\lambda ^{2}\varepsilon }.
\end{eqnarray}%
To look for a simple solution for $G_{\mu \nu }$; we choose $a$
and $b$ to be constants. Then choosing $a\gg \varepsilon $ and
$-\varepsilon <\lambda <\varepsilon $ gives the energy momentum
tensor:
\begin{eqnarray}
p_{space}=T_{11}=T_{22}=T_{33}&=&-\frac{4}{8\pi Ga_{0}}\\
-\rho =T_{44}&=&-\frac{6}{8\pi Ga_{0}^{2}}\\
p_{internal}=T_{55}=T_{66}&=&-\frac{3}{8\pi Ga_{0}^{2}} ,
\end{eqnarray}%
where $a_{0}=\varepsilon a$, $\rho $ is the energy density,
$p_{space}$ is
the pressure and $p_{internal}$ is the presure in internal space \cite%
{müller}$.$ All nondiagonal components of $T_{\mu \nu }$ are zero.
The internal and external spaces are both isotropic. This
implies the equations of state
\begin{eqnarray}
p_{space}&=&-\frac{2}{3}\rho , \\  \label{3}
p_{internal}&=&-\frac{1}{2}\rho .  \label{4}
\end{eqnarray}%
Here (\ref{3}) denotes the inflationary equation of state for a
space dominated with domain walls or membranes and (\ref{4})
interpreted as internal space dominated with strings. In general,
in a $d$ dimensional space, two dimensional relativistic
structures (i.e. walls or membranes) dominating a universe give
rise to the equation of state $p=-\frac{2}{d}\rho $ and one
dimensional relativistic structures (i.e. strings) give rise to
the equation of state $p=-\frac{1}{d}\rho .$ In Kaluza-Klein
terminology the name internal refers to the dimensions of space
which cannot be (yet) observed due to their microscopic extent. In
such an interpretation for the
metric in (\ref{2}), $\sigma $ and $\lambda $ should have microscopic extent %
\cite{kaluza}.



\begin{thebibliography}{99}
\bibitem{linde} Robertson, H.P. 1935. \textit{Proc. N.A.S. }15, 822,

\bibitem{walker} Walker, A.G. 1936. \textit{Proc. London Math. Soc. 42, 90,}

\bibitem{friedmann} Friedmann, A. 1922. \textit{Z Phys. }10, 377.

\bibitem{finsler} Finsler, P. 1918. \textit{\"{U}ber Kurven und Fl\"{a}chen
in Allgemeinen R\"{a}umen}, Doctoral Thesis, G\"{o}ttingen
University.

\bibitem{bao} Bao, D., Chern, S,-S., and Shen, Z. (Eds.). \textit{Finsler
Geometry. }Providence, RI: Amer. Math. Soc., 1996.

\bibitem{rund} Rund H 1959 \textit{The Differential Geometry of Finsler
Space }Springer-Verlag Ohg., Berlin. G\"{o}ttingen. Heidelberg.

\bibitem{godel} G\"{o}del K. 1949 \textit{Rev. Mod. Phys. }21, 447.

\bibitem{reboucas} Reboucas M J and Tiomno J 1982 \textit{Physical Rewiew D
28, 6, 1251-1264.}

\bibitem{taub1} Eguchi, T., Gilkey, P., Hanson, A. 1980 \textit{Phys. Rep. }%
66, 213-393.

\bibitem{gibbons} Gibbons G.W and Hawking S.W. 1979 \textit{Commun. Math.
Phys. }66, 29.

\bibitem{nergiz} Nergiz S. and Sa\c{c}l\i o\u{g}lu C. 1996 \textit{Physical
Rewiev D }53, 2240-2243.

\bibitem{müller} M\"{u}ller F.-Hoissen 1986 \textit{Class. Quant. Grav. }3,
665.

\bibitem{kaluza} Kaluza T. 1921 \textit{Sitzungsber. Presuss. Akad. Wiss.
Phys. Math. Kl. }966.
\end{thebibliography}
\end{document}